\documentclass[prl,a4paper,preprint,showpacs,byrevtex,onecolumn,showkeys]{revtex4}
\usepackage{graphicx}
\usepackage{amsmath}
\usepackage{array}
\usepackage{bm}
\usepackage{textcomp}

\begin{document}

\title{Excited flux tube from $q\bar q g$ hybrid mesons}

\author{Fabien \surname{Buisseret\, $^a$}}
\email[E-mail: ]{fabien.buisseret@umh.ac.be}
\author{Vincent \surname{Mathieu\, $^a$}}
\email[E-mail: ]{vincent.mathieu@umh.ac.be}
\author{Claude \surname{Semay\, $^a$}}
\email[E-mail: ]{claude.semay@umh.ac.be} 
\author{Bernard \surname{Silvestre-Brac\, $^b$}}
\email[E-mail: ]{silvestre@lpsc.in2p3.fr} 
\affiliation{$^a$Groupe de Physique Nucl\'{e}aire Th\'{e}orique, Universit\'{e} de Mons-Hainaut, Acad\'{e}mie universitaire Wallonie-Bruxelles, Place du Parc 20, BE-7000 Mons, Belgium.\\
$^b$Laboratoire de Physique Subatomique et de Cosmologie, Avenue des Martyrs 53, FR-38026 Grenoble-Cedex, France.}

\date{\today}

\begin{abstract}
In the framework of quark models, hybrid mesons are either seen as two-body $q\bar q$ systems with an excited flux tube connecting the quark to the antiquark or as three-body $q\bar q g$ systems including a constituent gluon. In this work we show that, starting from the three-body wave function of the $q\bar q g$ hybrid meson in which the gluonic degrees of freedom are averaged, the excited flux tube picture emerges as an equivalent $q\bar q$ potential. This equivalence between the excited flux tube and the constituent gluon approach is confirmed for heavy hybrid mesons but, for the first time, it is shown to hold in the light sector too, provided the contribution of the quark dynamics is correctly taken into account.
\end{abstract}

\maketitle 

The study of hybrid mesons is an active domain in theoretical and in experimental particle physics. From a theoretical point of view, they are interpreted as mesons in which the color field is in an excited state. Numerous lattice QCD calculations have been devoted to hybrid mesons \cite{Nel02,Juge}, as well as many studies involving effective models. In particular, within the framework of quark models, there are two main approaches. In the first one, the quark and the antiquark are linked by a string, or flux tube, which is responsible for the confinement. In this stringy picture, it is possible for the flux tube to fluctuate, and thus to be in an excited state \cite{Allen:1998wp,luscher}. The second approach assumes that the hybrid meson is a three-body system formed of a quark, an antiquark, and a constituent gluon. Two straight strings then link the gluon to the quark and to the antiquark. This picture has been firstly studied in Refs.~\cite{constg}, but also in more recent works \cite{Szczepaniak:2006nx,Mathieu:2005wc,Buisseret:2006sz}. 

It was suggested in Ref.~\cite{Buisseret:2006sz} that, in the static quark limit, the constituent gluon picture is equivalent to the excited flux tube one, the total energy of the constituent gluon being equal to the energy contained in the excited string. These results were extended in Ref.~\cite{Buisseret:2006wc}, where the dynamics of the quarks has been taken into account. For further developments, it is useful to briefly sum up the key points of this last reference.

Assuming the Casimir scaling hypothesis, it can be shown that the flux tubes in a $q\bar q g$ system are two straight strings linking the gluon to the quark and to the antiquark \cite{Mathieu:2005wc}, in agreement with Refs.~\cite{constg}. In this case, taking only the confining interaction into account, we can write a spinless Salpeter Hamiltonian for the system, 
\begin{equation}\label{mainH}
H_{3b}=\sum_{i=q,\bar q,g}\sqrt{\bm p^2_i+m^2_i}+\sum_{j=q,\bar q}a|\bm x_j-\bm x_g|,
\end{equation}
with $m_g=0$. The three-body eigenequation
\begin{equation}\label{eig1}
H_{3b}\Psi_{3b}(\bm r,\bm y)=M_{3b}\Psi_{3b}(\bm r,\bm y)
\end{equation}
can be analytically solved by using the auxiliary field technique if the quark and the antiquark are of the same mass. We will only focus on this case in the following. It is worth noting that $\bm r=\bm x_q-\bm x_{\bar q}$ is the quark-antiquark separation, and that $\bm y$ is the second relative variable, directly linked to the gluon position. Various approximations are necessary to perform this resolution, and lead to eigenfunctions which are separable, i.e.
\begin{equation}\label{separa}
\Psi_{3b}(\bm r,\bm y)=A(\bm r) B(\bm y).
\end{equation}
We stress that this exact separability is only an artifact of the auxiliary field technique. 

It can be shown that, by dropping the ``gluonic" part $B(\bm y)$, the quark-antiquark wave function $A(\bm r)\equiv\Psi_{2b}(\bm r)$ satisfies the eigenequation
\begin{subequations}\label{effpotg}
\begin{equation}
H_{2b}\Psi_{2b}(\bm r)=M_{3b}\Psi_{2b}(\bm r),
\end{equation}
where $H_{2b}$ is the two-body spinless Salpeter Hamiltonian
\begin{equation}
H_{2b}=\sum_{i=q,\bar q}\sqrt{\bm p^2_i+m^2_i}+V_{q\bar q}(r),
\end{equation}
\end{subequations}
with $V_{q\bar q}$ the equivalent two-body potential.

For heavy quarks, this potential has the form \cite{Buisseret:2006wc}
\begin{equation}\label{Vqq}
V^h_{q\bar q}(r)=\sqrt{\sigma^2r^2+2\pi\sigma({\cal N}+3/2)},
\end{equation}
where $\sigma$ is a new string tension. In this potential, ${\cal N}=2n_y+\ell_y$, with $n_y$ and $\ell_y$ the radial quantum number and the orbital angular momentum with respect to the variable $\bm y$. Consequently, ${\cal N}$ defines the gluon state, and the equivalent potential depends on this quantum number. Actually, formula~(\ref{Vqq}) is nothing but the energy of an excited string of length $r$ \cite{arvi}, whose square zero point energy is given by $3\pi\sigma$ (for ${\cal N}=0$). It is different of the generally accepted value in string theory, which is $-2\pi a(D-2)/24$, with $D$ the dimension of space (see for example Ref.~\cite[p.~231]{str}). Together with $D=26$, it ensures that the Lorentz invariance is still present at the quantum level. However, the string we are dealing with is an effective one at $D=4$, simulating the confining interaction for an excited color field. In that sense, our nonstandard value is more relevant for the study of hybrid mesons, since $3\pi\sigma$ is actually equal to the square zero point energy of the gluon and the two strings in the $q\bar q g$ system \cite{Buisseret:2006wc}. It is also important to notice that the string tension $\sigma$ is not necessarily equal to the string tension $a$, since the excited string is an effective object emerging from the gluon-plus-string system. 

In the limit where the quarks are massless, computations are more complex, and only an asymptotic approximate form can be computed for the equivalent two-body potential, that is \cite{Buisseret:2006wc}
\begin{equation}\label{Vqql}
V^l_{q\bar q}(r \gg \sigma^{-1/2})\approx \sigma r+\frac{4}{r} ({\cal N}+3/2).
\end{equation}

The purpose of this work is to extend the results of Refs.~\cite{Buisseret:2006sz,Buisseret:2006wc} by performing an accurate numerical resolution of the eigenequation~(\ref{eig1}) followed by an accurate numerical inversion of the eigenequation~(\ref{effpotg}). It is then possible to compute the equivalent potential $V_{q\bar q}$ for the $q\bar q$ pair contained in the $q\bar q g$ system from the corresponding internal $q\bar q$ wave function. These results, obtained without the approximations of the auxiliary field method, will be computed for hybrid mesons formed of heavy as well as for light quarks of the same flavor, and then compared to Eqs.~(\ref{Vqq}) and (\ref{Vqql}). The procedure is the following:

Firstly, the mass $M_{3b}$ and the three-body wave function $\Psi_{3b}(\bm r, \bm y)$ have to be computed. For the lowest lying state, the wave function reads
\begin{align}\label{decompose}
\Psi_{3b}(\bm r, \bm y)=&\left[\left[\bm 3,\bar{ \bm 3}\right]^{\bm8},\bm 8\right]^{\bm 1}\otimes\left[\left[i,i\right]^{I},0\right]^{I} \nonumber \\
&\otimes\left[\left[1/2,1/2\right]^{S_{q\bar q}},1\right]^{S}\otimes\Phi_{L=0}(\bm r,\bm y),
\end{align}
where the color, isospin, spin, and space functions have been explicitly written. The color function is unique since the hybrid meson is in a color singlet, and the isospin function is trivial. The spin functions are unambiguously given when $S=0$ or $2$ since $S_{q\bar q}=1$. Furthermore, as the C-parity, $C=(-)^{S_{q\bar q}+1}$, is a good quantum number \cite{constg}, there is no coupling between $S=1$ states with either $S_{q\bar q}=0$ or $1$. The spin function is thus always unique. As we deal with the ground state of Hamiltonian~(\ref{mainH}), only the $L=0$ spatial wave function has to be computed. It means that the parity $P$ of the state is always positive \cite{constg}. For this space wave function, Gaussian trial functions of the following form are used
\begin{equation}\label{trial}
\Phi_{L=0}(\bm r, \bm y)=\sum^N_{i=1} \alpha_i\exp\left(-a_i\, r^2 - b_i\, y^2-2\, c_i\, \bm r\cdot\bm y\right),
\end{equation}
where the variational parameters $a_i$, $b_i$, and $c_i$ have to be determined by a minimization on the mass $M_{3b}$ (see Ref.~\cite{suzu} for more details on this method). Calculations are simplified if we set $c_i=0$ in Eq.~(\ref{trial}), which means that the orbital angular momenta relative to the Jacobi coordinates $\bm r$ and $\bm y$ are both equal to zero. It can be checked that this has a little influence -- around $1\%$ -- on the ground state mass $M_{3b}$. As Hamiltonian $H_{3b}$ is spin- and isospin-independent, the states with $S=0,1,2$ and $I=0,1$ (for light quarks) are degenerate. Consequently, the $0^{++},1^{++},1^{+-}$, and $2^{++}$ states are degenerate ground states of our model. 


Secondly, the two-body wave function $\Psi_{2b}(\bm r)$ has to be computed from $\Psi_{3b}(\bm r, \bm y)$ (up to an irrelevant normalization factor). Due to good $C$ and $P$ quantum numbers, the color, isospin, and spin functions of the $q\bar q$ pair are uniquely determined from $\Psi_{3b}(\bm r, \bm y)$. A general way of averaging the space gluonic degrees of freedom is to solve the following equation
\begin{equation}\label{presc_general}
\Psi^n_{2b}(\bm r)=\int f(\bm y)\Phi^n_{L=0}(\bm r,\bm y)d\bm y, 
\end{equation}
where $n$ is an arbitrary exponent, and $f(\bm y)$ an arbitrary weight function. If $\Phi^n_{L=0}(\bm r,\bm y)$ were separable as in Eq.~(\ref{separa}), any choice of $f(\bm y)$ and $n$ would give identical results. This is not a priori the case for the trial wave function~(\ref{trial}). We use in the following the two simplest prescriptions, both characterized by $f(\bm y)=1$. We call the first one the linear prescription, with $n=1$ and 
\begin{equation}
\Psi_{2b}(r)=\sum^N_{i=1}\alpha_i\, \left(\frac{\pi}{b_i}\right)^{3/2}\exp(-a_i r^2),
\end{equation}
where $r$ is the quark-antiquark distance. This prescription is the most straightforward way to remove the space gluonic part from the three-body wave function. The second prescription we suggest is called the quadratic prescription, for which $n=2$,  
\begin{equation}
\Psi^2_{2b}(r)=\sum^N_{i,j=1}\alpha_i\, \alpha_j\, \left(\frac{\pi}{b_i+b_j}\right)^{3/2}\exp\left[-(a_i+a_j) r^2\right].
\end{equation}
It has the interesting feature that $\int \Psi^2_{2b}(r)\, dr=1$ is a priori satisfied when $\Psi_{3b}$ has a unit normalization. In general, when taking the square root, the sign has to be chosen such that $\Psi_{2b}(r)$ is smooth, but no such choice has to be done here since we deal with the ground state wave function. The regularized $q\bar q$ wave function is then given by $u(r)=r\, \Psi_{2b}(r)$.

Thirdly, knowing the two-body wave function $u(r)$, the mass $M_{3b}$, and the quark masses (same values as in Hamiltonian $H_{3b}$), the equivalent potential $V_{q\bar q}(r)$ satisfying Eqs.~(\ref{effpotg}) can be computed. We gave in Ref.~\cite{lag1} a procedure, relying on the Lagrange mesh method \cite{baye86,sem01}, to make such computations with a semirelativistic kinematics. This procedure has been shown to be accurate up to $3\%$ in the computation of the equivalent potential, and has already been applied to study the effective potential between two gluons \cite{glue2}. 


\begin{figure}[ht]
\begin{center}
\includegraphics*[width=8cm]{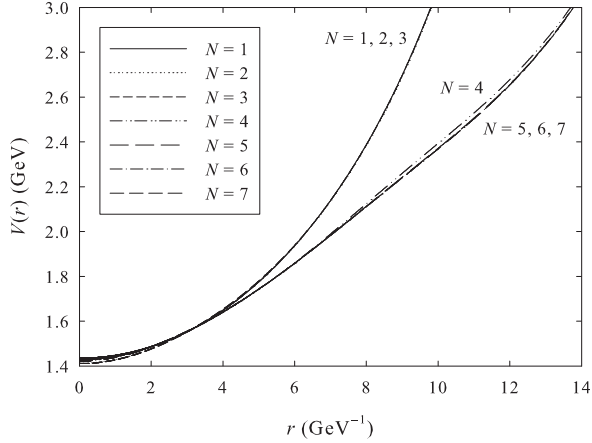}
\end{center}
\caption{Evolution of the equivalent $c\bar c$ potential for different values of $N$. The linear prescription and the parameters of Table~\ref{params} are used.}
\label{fig:ccg1}
\end{figure}

The procedure we have described can now be applied to a physical system. Let us begin with the $c\bar cg$ hybrid, for which we use the parameters of Ref.~\cite{Mathieu:2005wc}: $m_c=1.3$ GeV and $a=0.175$ GeV$^2$. These values are fitted so that the mass of the $J/\Psi$ meson is correctly reproduced as well as the Regge slope of the $\rho$ meson family. We begin by showing the evolution of the equivalent potential with the number of Gaussian functions in Fig.~\ref{fig:ccg1}. It appears that the shape of the potential does not change anymore when $N\geq6$. Moreover for $N=7$, the error on the $c\bar c g$ mass is less than $1$ MeV, which is a very satisfactory precision. Although we only plotted the effective potential computed with the linear prescription, we checked that the same conclusions hold for the quadratic one but also for other choices.

\begin{table}[ht]
\caption{Numerical values for the different parameters involved in our computations of $q\bar q g$ systems. $a=0.175$ GeV$^2$, $m_g=0$, $m_n$ and $m_c$ are taken from Ref.~\cite[Table 2]{Mathieu:2005wc}. The symbol $n$ is used for both the $u$ and $d$ quarks.}
\begin{tabular}{lcccc}\label{params}
\\ \hline\hline
Quark & $b$   & $c$   & $s$   & $n$\\
\hline
$m_q$ (GeV)       & 4.567 & 1.300 & 0.180 & 0.000\\
$M_{3b}$ (GeV)    &10.665 & 4.323 & 2.516 & 2.421\\
$\sigma$ (GeV$^2$)& 0.200 & 0.200 & 0.207 & 0.207\\
$W_q$ (GeV$^2$)   & 0.000 & 0.160 & 0.460 & 0.530\\
$\mu_q$ (GeV)     & 4.757 & 1.409 & 0.435 & 0.374\\	
\hline\hline	
\end{tabular}
\end{table}

\begin{figure}[ht]
\begin{center}
\includegraphics*[width=8cm]{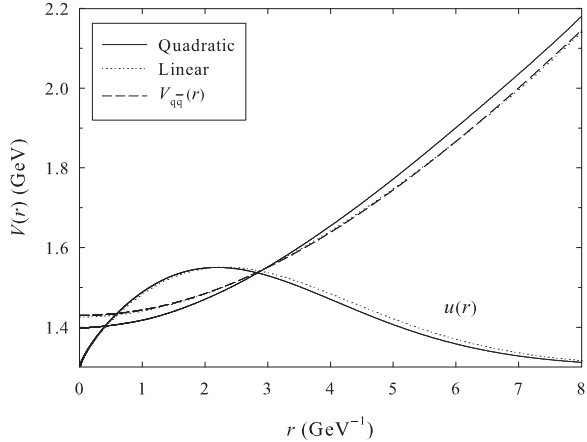}
\end{center}
\caption{Effective potentials and regularized wave functions (with an arbitrary normalization) computed with $N=7$ for the $c\bar c g$ system with the quadratic prescription (solid line) and the linear one (dotted line). Potential~(\ref{vmod}) is plotted for ${\cal N}=0$ (dashed line); it is practically indistinguishable from the result of the linear prescription on this figure. The various parameters are summed up in Table~\ref{params}.}
\label{fig:ccg2}
\end{figure}

The equivalent $c\bar c$ potential is then computed with $N=7$, with the linear and quadratic prescriptions, and plotted in Fig.~\ref{fig:ccg2}. It can be seen that both prescriptions roughly lead to the same potential. Furthermore, we actually checked that the equivalent potential is nearly independent of the prescription which is used. This means that the dynamics of the gluon is quasi-decoupled from the quark-antiquark pair in the $c\bar c g$ system, and more generally in $q\bar q g$ systems. The equivalent potential can be fitted by the following generalization of potential~(\ref{Vqq})
\begin{equation}\label{vmod}
V_{q\bar q}(r)=\sqrt{\sigma^2r^2+2\pi\sigma({\cal N}+3/2)+W_q}, 
\end{equation}
where ${\cal N}=0$. It is checked that the introduction of a constant $W_q\neq0$ in potential~(\ref{vmod}) is necessary to recover the numerical data (see Table~\ref{params}).

\begin{figure}[ht]
\begin{center}
\includegraphics*[width=8cm]{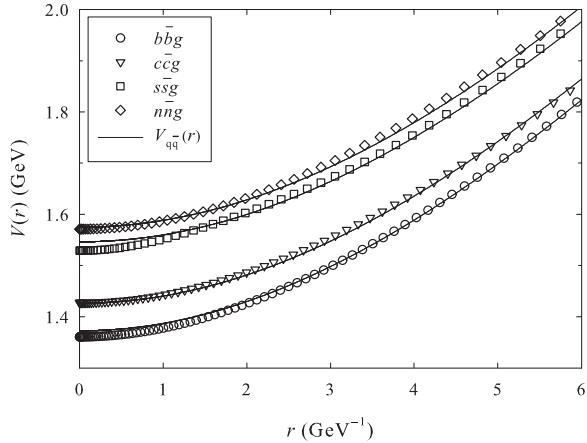}
\end{center}
\caption{Effective potentials computed with $N=7$ and the linear prescription for the different $q\bar q g$ systems (symbols). Potential~(\ref{vmod}) is plotted for ${\cal N}=0$ (solid lines). The various parameters are summed up in Table~\ref{params}.}
\label{fig:ccg3}
\end{figure} 

Very good descriptions of mesons and baryons, even light ones, are obtained within the framework of potential models. It has been recently shown that such models give $gg$ and $ggg$ glueball spectra very similar to results predicted by lattice QCD works \cite{glue}. So we can assume that the procedure developed in this paper to study hybrid mesons remains valid for light quarks. We thus computed the effective potential $V_{q\bar q}$ for all ground state $q\bar q g$ systems. Results are presented in Fig.~\ref{fig:ccg3}. Let us note that the masses of the $s$ and $b$ quarks were fitted on the masses of the $\phi(1020)$ and of the $\Upsilon(1S)$ by using the model of Ref.~\cite{Mathieu:2005wc}. It appears from a fit that the equivalent potentials are all compatible with the form~(\ref{vmod}), with $\sigma\approx 0.200\pm0.010$~GeV$^2$, and a value of $W_q$ which increases as the quarks become lighter (see Table~\ref{params}). 

As we mentioned before, $\sigma\neq a$ since the emergent excited string has not to share the same string tension than the two initial strings in the $q\bar q g$ system. Such a feature could not be found in the previous study of Ref.~\cite{Buisseret:2006wc} due to the approximate method used. It has been shown in Ref.~\cite{neste} that the energy of a string in its ground state is affected by the presence of dynamical particles at its ends. In the same way, we argue that $W_q$ could be a modification of the square zero point energy of the two strings because of quark dynamics. A first hint is that $W_b=0$ is compatible with the data. This corresponds to the static limit of potential~(\ref{Vqq}), in agreement with lattice QCD \cite{Juge}. For the first time however, we are able to show that the excited string picture also holds for light quarks, but with a nonzero value for $W_{q}$. A comparison of potential~(\ref{vmod}) with the approximate result~(\ref{Vqql}) for $r \gg \sigma^{-1/2}$ leads, for ${\cal N}=0$, to $W_n\approx (4-\pi)\, 3\, \sigma\approx 0.52\ \rm{GeV}^2$ in the limit of massless quarks, in agreement with the fitted value. Moreover, defining the dynamical mass of the quark as $\mu_q=\left\langle \Psi_{3b}\right| \sqrt{\bm p^2_q + m^2_q}\left|\Psi_{3b}\right\rangle$ \cite{glue}, it can be checked that $W_q\approx0.2/\mu_q$ reproduces very well the fitted values given in Table~\ref{params}. Consequently, $W_q$ is directly related to the quark dynamics at the end of the strings.

In conclusion, we have shown that, in the framework of a potential model with spinless constituent quarks and gluons, it is possible to compute the dominant interaction between a quark and an antiquark in a hybrid meson from a $q\bar q g$ system. The potential obtained is in good agreement with results predicted by lattice QCD calculations for heavy quarks \cite{Juge}, provided the string tension is slightly greater in hybrid mesons than in ordinary mesons. Moreover, up to our knowledge, we predict, for the first time, that the dominant quark-antiquark interaction in a hybrid meson depends on the quark flavor just by a square shift in energy, due to the quark dynamics. 

Since our method can easily be generalized to spin-dependent Hamiltonians, we think that these results open the way for a deeper understanding of the fine structure of the QCD string spectrum: By adding proper short-range interactions \cite{eqpot}, the equivalent potentials will also depend on $S_{q\bar q}$ through spin-orbit or spin-spin couplings. These couplings could explain the short-range splittings of the potentials as they are observed in lattice QCD \cite{Juge}. We leave such studies for future works. 

\acknowledgments 
C. S. and F. B. thank the FNRS for financial support. V. M. thanks the IISN for financial support.

\end{document}